\documentclass[a4paper]{article}

\usepackage{hyperref}
\usepackage{cite}

\usepackage[justification=centering]{caption}
\usepackage[labelformat=simple]{subcaption}

\usepackage{pifont}% http://ctan.org/pkg/pifont
\newcommand{\cmark}{\ding{51}}%
\newcommand{\xmark}{\ding{55}}%

\usepackage{INTERSPEECH2020}

\title{The NU Voice Conversion System for the Voice Conversion Challenge 2020: On the Effectiveness of Sequence-to-sequence Models and Autoregressive Neural Vocoders}
\name{Wen-Chin Huang$^{\star}$\thanks{$\star$ Equal contribution.}, Patrick Lumban Tobing$^{\star}$, Yi-Chiao Wu$^{\star}$, Kazuhiro Kobayashi$^{\star}$, Tomoki Toda}

\address{Nagoya University, Japan}
\email{
    \{wen.chinhuang, patrick.lumbantobing, yichiao.wu\}@g.sp.m.is.nagoya-u.ac.jp, kobayashi.kazuhiro@g.sp.m.is.nagoya-u.ac.jp, tomoki@icts.nagoya-u.ac.jp
}

\begin{document}

\maketitle
\begin{abstract}
In this paper, we present the voice conversion (VC) systems developed at Nagoya University (NU) for the Voice Conversion Challenge 2020 (VCC2020). We aim to determine the effectiveness of two recent significant technologies in VC: sequence-to-sequence (seq2seq) models and autoregressive (AR) neural vocoders. Two respective systems were developed for the two tasks in the challenge: for task 1, we adopted the Voice Transformer Network, a Transformer-based seq2seq VC model, and extended it with synthetic parallel data to tackle nonparallel data; for task 2, we used the frame-based cyclic variational autoencoder (CycleVAE) to model the spectral features of a speech waveform and the AR WaveNet vocoder with additional fine-tuning. By comparing with the baseline systems, we confirmed that the seq2seq modeling can improve the conversion similarity and that the use of AR vocoders can improve the naturalness of the converted speech.

\end{abstract}

\section{Introduction}

Voice conversion (VC) aims to convert in speech from a source to that of a target without changing the linguistic content \cite{VC, GMM-VC}. The voice conversion challenge\footnote{\url{http://www.vc-challenge.org/}} (VCC) aims to better understand advances in VC techniques. This year \cite{vcc2020summary}, two tasks were considered: the first task was \textit{intra-lingual semiparallel} VC, where only a small subset of the training set was parallel, with the rest being nonparallel; the second task was \textit{cross-lingual} VC, where the training set of the source speaker is different from that uttered by the target speaker in language and content, thus nonparallel in nature. In conversion, the source speaker's voice in the source language is converted as if it was the target speaker's voice.

In recent years, deep learning has been a game changer in many research fields, and VC is no exception. The theoretically unlimited expressive power of deep neural networks (DNNs) makes it possible to model various complex characteristics that are essential to the performance of VC systems, such as the high resolution nature of speech signals and the conversion of prosody and speaking rates. These characteristics have been addressed by two epoch-making technologies: sequence-to-sequence (seq2seq) models and neural vocoders.

First, as most VC studies have focused on frame-by-frame conversion models, i.e., the converted speech and the source speech are always of the same length, the modeling of speaking rate is largely restricted, and so are other prosody-related factors such as the F0 contour. Seq2seq models \cite{S2S}, which are often equipped with an attention mechanism \cite{S2S-NMT-Bah, S2S-NMT-Luong} to implicitly learn the alignment between the source and output sequences and generate outputs of various lengths and capture long-term dependences, are therefore suitable for converting prosody in VC. Since the suprasegmental characteristics of F0 and duration patterns well handled in seq2seq VC models are closely correlated with the speaker identity, it has been shown that seq2seq VC models can outperform conventional frame-wise VC systems, especially in terms of conversion similarity \cite{ATT-S2S-VC, S2S-iFLYTEK-VC}. 

On the other hand, neural-based speech generation models~\cite{wnv, wnv-voc, shallow-wnv, samplernn, wavernn, lpcnet, clarinet, waveglow, flowavenet, nsf_2019, parallel-wavegan, melgan, gantts, qppwg} have been proposed to directly model speech waveforms, to overcome the naturalness degradation caused by the loss of phase and temporal details from the oversimplified assumptions in conventional parametric-based vocoders~\cite{straight, world}. Although autoregressive (AR) models~\cite{wnv, wnv-voc, shallow-wnv, samplernn, wavernn, lpcnet} achieve marked high-fidelity speech generation, the AR mechanism greatly limits the generative speed or model complexity. That is, to achieve real-time generation, a compact AR model with specific prior knowledge~\cite{wavernn, lpcnet} is required. On the other hand, on the basis of the parallel computing advantage of convolution neural networks (CNNs), many non-AR models such as flow-based~\cite{clarinet, waveglow, flowavenet}, source-filter-based~\cite{nsf_2019}, and GAN-based~\cite{parallel-wavegan, melgan, gantts, qppwg} models have been proposed. However, the naturalness of VC speech generated by these non-AR models has not been comprehensively evaluated and compared with AR models. Therefore, we provide more insight via the NU VC system in this paper.

Here, we describe our submission to the VCC2020. For task 1, we extended our previously proposed Transformer-based seq2seq VC framework, the Voice Transformer Network (VTN) \cite{VTN, VTN-TASLP} with synthetic parallel data, where we used text-to-speech (TTS) systems to generate synthetic data for parallel learning. For task 2, we used the frame-based cyclic variational autoencoder (CycleVAE) \cite{cyclevae} framework to model the spectral features of the speech waveform and the AR WaveNet vocoder \cite{wnv,wnv-voc,shallow-wnv} as the waveform generator with additional fine-tuning. We aim to answer the following research questions:
\begin{itemize}
    \item In what aspects can seq2seq VC models outperform frame-based VC?
    \item How can VC systems benefit from AR neural vocoders over non-AR neural vocoders?
\end{itemize}

\section{Task 1: VTN with synthetic parallel data}

\subsection{Motivation}
\label{ssec:motivation}

\begin{figure*}[t]
  \centering
  \includegraphics[width=\linewidth]{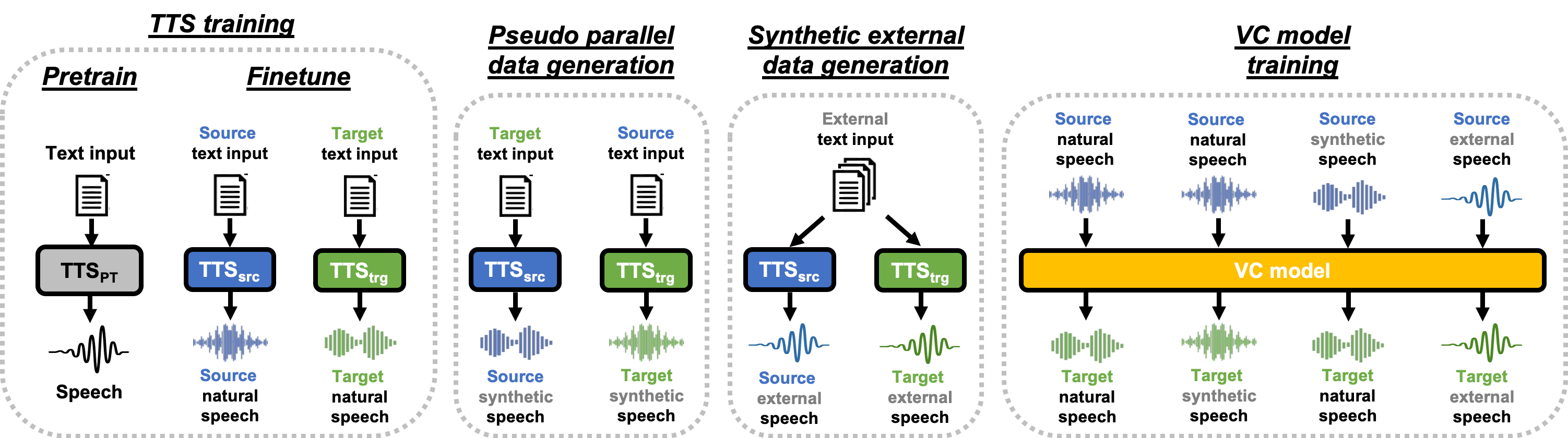}
  \caption{Overview of the training procedure of task 1 system. The speech utterances generated by a TTS system are shown in grey.}
  \label{fig:task1-overview}
\end{figure*}

Most seq2seq VC models are data-hungry, requiring over 1 h of data to generalize well. Although we previously proposed a TTS-based pretraining technique that can generalize seq2seq VC models to only 5 min of training data \cite{VTN, VTN-TASLP}, a major limitation of most existing seq2seq VC models, including ours, is that they can only achieve \textit{parallel} VC, i.e., a requirement of a parallel corpus consisting of pairs of speech samples with identical linguistic contents uttered by both source and target speakers. The application of current seq2seq VC models to a semiparallel setting such as task 1 in VCC2020 is therefore not straightforward, since among the 70 utterances in total, only 20 are parallel and the remaining 50 are not.

Inspired by \cite{S2S-parrotron-VC}, we utilized TTS models to generate synthetic parallel data for VTN training. Considering the corpus in task 1 being semiparallel, there are three types of speech to use:
\begin{enumerate}
    \item \textbf{Natural speech}: the utterances in the corpus.
    \item \textbf{Synthetic pseudo-parallel speech}: The synthetic target/source counterpart of the natural speech is generated by inputting the transcription of a source/target natural speech to a target/source TTS system.
    \item \textbf{Synthetic external speech}: A synthetic pair of utterances with identical content is generated by inputting a set of external text to the source and target TTS systems.
\end{enumerate}
As a result, there are four types of parallel data pairs available for training the VC model in total:
\begin{enumerate}
    \item \textless source natural, target natural\textgreater
    \item \textless source synthetic pseudo-parallel, target natural\textgreater
    \item \textless source natural, target synthetic pseudo-parallel\textgreater
    \item \textless source synthetic external, target synthetic external\textgreater
\end{enumerate}

\subsection{System Overview}

In this section, we describe the training and conversion procedure in detail. Figure~\ref{fig:task1-overview} illustrates the training procedure of our task 1 system. 

\subsubsection{TTS training}
\label{sssec:tts-training}

To generate a synthetic speech, TTS systems for the source and target speakers need to be trained first. We adopted a seq2seq-based TTS model considering the success of seq2seq models in the field of TTS. However, the size of the training set of each target speaker in VCC2020 is too limited to train seq2seq TTS models from scratch. In light of this, we employed a pretraining\textendash finetuning scheme that first pretrains on a large TTS dataset followed by fine-tuning on the limited source or target speaker dataset. This allowed us to successfully train on even 5 min of data.

\subsubsection{Pseudo-parallel data generation}

For the source/target natural utterances in the corpus, we can generate the  target/source counterpart to form a \textless natural, synthetic\textgreater\ data pair. For instance, by simply passing the transcription of a source utterance to the target TTS system mentioned in Section~\ref{sssec:tts-training}, the output will be of the identity of the target speaker but with the same content as that of the source utterance. This utterance can therefore be used together with the source natural utterance to form a parallel data pair.

\subsubsection{Synthetic external data generation}

Even with pseudo-parallel data, the total number of valid parallel data pairs is still only 120. For data augmentation, we used an external set of text data as input and generated synthetic voices using the source and target TTS systems. Since we pass the same contents, this set of synthetic external data is parallel in nature and can be used to train the seq2seq VC model. In our initial experiments, we found that including such data can improve the intelligibility of the converted speech.

\subsubsection{VC model training and conversion process}

Using the four types of parallel data pairs described in Section~\ref{ssec:motivation}, we can extract acoustic features from them as inputs and outputs to train the seq2seq VC model in a parallel manner as usual. In the  conversion phase, given a source speech utterance, we first extract the acoustic feature and pass it to the seq2seq VC model to obtain the converted acoustic features. A neural vocoder is then used to synthesize the final converted waveform.

\subsection{Implementation}

The seq2seq VC model was of the same architecture as described in \cite{VTN-TASLP}, and we used the official implementation\footnote{\url{https://github.com/espnet/espnet/tree/master/egs/arctic/vc1}}. For TTS model training, we directly used the recipes provided in the open-source implementation of the VCC2020 seq2seq baseline\footnote{\url{https://github.com/espnet/espnet/tree/master/egs/vcc20}}. The external text that we used was from the CMU ARCTIC dataset \cite{CMU-Arctic} which contained 1132 utterances. For the neural vocoder, we used the non-AR, faster than real-time Parallel WaveGAN (PWG) \cite{parallel-wavegan}, and we directly used the pretrained models from the VCC2020 seq2seq baseline.

\section{Task 2: CycleVAE and WaveNet vocoder}

In this section, we describe the NU system for Task 2 using the CycleVAE-based \cite{cyclevae} spectral model and WaveNet-based \cite{shallow-wnv} vocoder. Compared with that in the official baseline systems \cite{vcc2020-baseline-cyclevae}, which also utilizes the CycleVAE-based spectral model, we used the AR neural vocoder, i.e., WaveNet, instead of a non-AR neural vocoder, i.e., PWG. Furthermore, we also used more variations of speech data to support the development of the WaveNet vocoder for producing high-quality synthetic speech. Therefore, compared with the CycleVAE baseline system T16 \cite{vcc2020-baseline-cyclevae}, the differences are as follows: (1) the use of more speech data for training CycleVAE and neural vocoder models, and (2) the use of an AR-based neural vocoder instead of a non-AR one.

\subsection{CycleVAE-based spectral model}

CycleVAE \cite{cyclevae} is a frame-based nonparallel spectral modeling framework based on the variational autoencoder (VAE) \cite{vae} and a cycle-consistent approach that recycles converted spectra for generating cyclic-reconstructed spectra that can be utilized in the optimization. This technique has been found to be very effective, compared with only using reconstructed spectra in the optimization \cite{vaevc}, to improve both the latent space condition, i.e., to be more speaker-independent, and the accuracy of the converted spectra. In this work, we follow the CycleVAE architecture described in \cite{cyclevae}, where the differences are the use of a standard Laplacian prior, the use of only two cycles, and the use of a unified encoder\textendash decoder for many-to-many VC.

\subsection{WaveNet vocoder}

The WaveNet vocoder \cite{wnv,wnv-voc} is an AR neural vocoder that can produce synthetic speech waveforms with natural quality. It consists of a stack of dilated convolutional blocks to effectively capture the receptive field of waveform samples. In this work, we follow the architecture of the shallow WaveNet vocoder using softmax output as in \cite{shallow-wnv}, with an extension of the fine-tuning procedure to reduce the mismatches \cite{vc-wnv-ft} between naturally extracted spectra and converted spectra generated from a VC model, such as from CycleVAE.

\subsection{Training and conversion process}

\begin{figure}[!t]
  \centering
  \includegraphics[width=\linewidth]{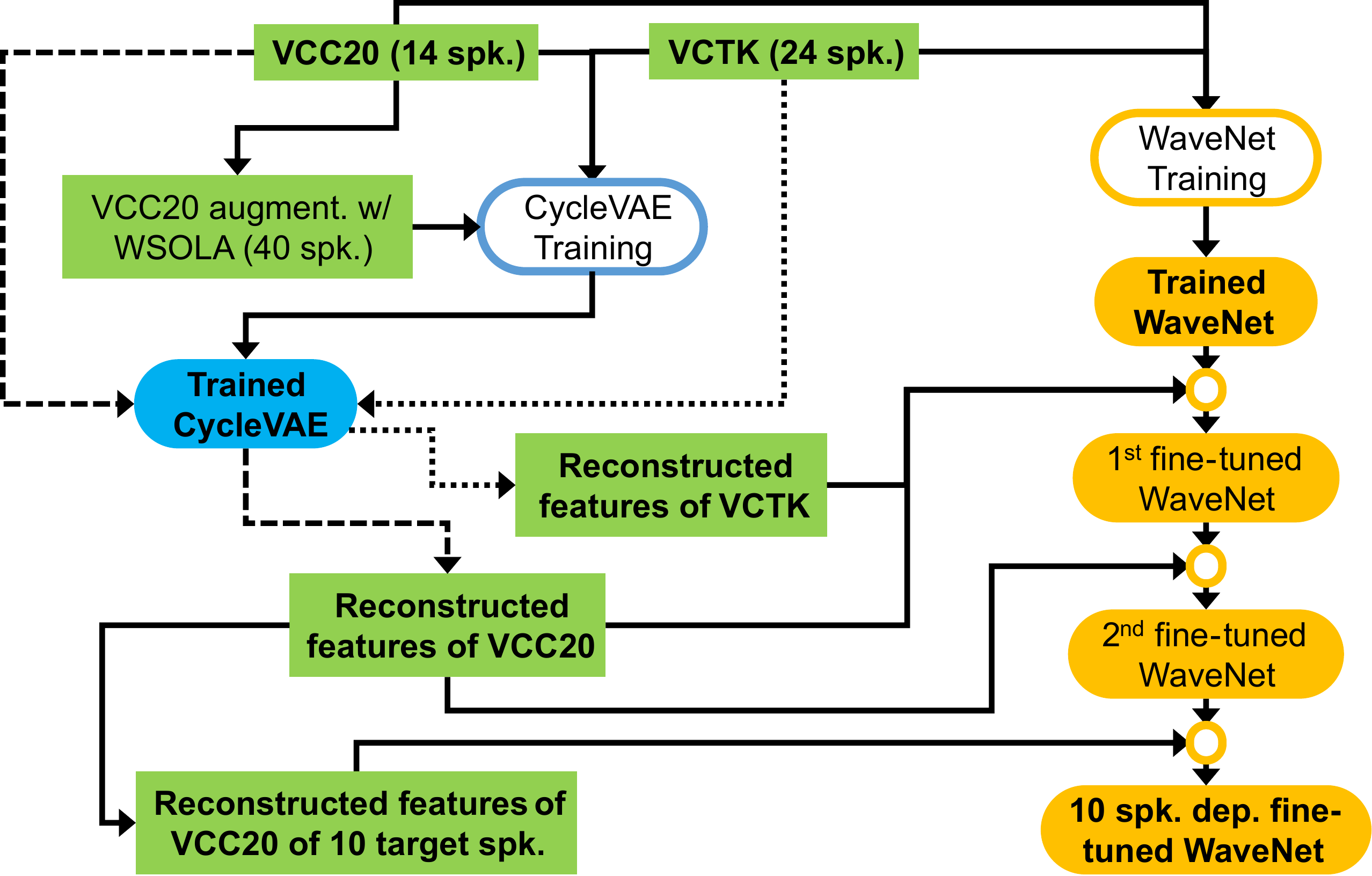}
  \caption{Development flow of CycleVAE-based spectral model and WaveNet vocoder for task 2 of NU system.}
  \label{fig:task2_train}
  \vspace*{-6mm}
\end{figure}

The development procedure for the CycleVAE and WaveNet vocoder in task 2 is shown in Fig.~\ref{fig:task2_train}. We used additional speech data of 24 speakers from the VCTK \cite{vctk} corpus, 12 males and 12 females, each with 315 utterances, to accompany the VCC 2020 dataset. To train the CycleVAE-based spectral model, we also performed data augmentation using waveform similarity and overlap add (WSOLA)-based F0 transformation \cite{wsola-f0} to produce modified speech waveforms from the speech waveforms available from the VCC 2020 dataset. Specifically, for each of the 10 target speakers, F0 ratios with respect to each of the four source speakers were used to generate modified speech waveforms producing 40 additional speakers.

On the other hand, in the training of the WaveNet vocoder, we carried out a four steps training/fine-tuning procedure to reduce the mismatches between natural and converted spectra. In the first step, we used natural speech features of the 38 speakers from VCTK and VCC 2020 datasets to train a natural multispeaker model. In the second step, we retrained the first model using speech features containing reconstructed spectra obtained from the CycleVAE model, i.e., from the 38 speakers. In the third step, we fine-tuned the second model using the generated speech features of only 14 speakers from the VCC 2020 dataset. Finally, we carried out the last fine-tuning using the generated features of only the target speakers to produce speaker-dependent WaveNet vocoders.

The above second training phase was performed in $\sim\!\!5.5$ days with NVIDIA Titan V with the number of optimization steps similar to that in the first training. The third and fourth phases were performed until the overfit condition was reached for the development set, which contained 10 utterances of the total 70 utterances for each speaker in the VCC 2020 dataset.

As the speech features, instead of the mel-spectrogram, we used WORLD-based \cite{world} features for both CycleVAE and WaveNet containing unvoiced/voiced (U/V) decisions, continuous log fundamental frequency (log-F0) values, aperiodicity parameters, and $49$-dimensional mel-cepstrum \cite{mel-cep} parameters including $0$-th power extracted from the WORLD-based spectral envelope. CycleVAE models only the estimation of mel-cepstrum parameters. In the conversion phase, we simply generated the converted spectra using the speaker code (one-hot vector) of the desired target speaker and the latent features of the source speaker as in \cite{cyclevae}. Then, linear transformation of log-F0 values \cite{GMM-VC} of the source speaker was performed to convert the pitch of the source to that of the target. Finally, to generate the converted speech waveform, we fed the converted speech features containing input U/V decisions, converted F0, input aperiodicity, and converted spectra to the WaveNet vocoder.

\begin{table*}[t]
	\centering
	\captionsetup{justification=centering}
	\caption{Subset of systems that participated in VCC2020.}
	
	\centering
	\begin{tabular}{ c c c c c }
		\toprule
		Team ID & Description & Conversion Model & Vocoder & AR \\  
		\midrule
		T10 & VCC2020 top system & PPG $\to$ \textbf{seq2seq} $\to$ mel filterbanks & WaveNet & \cmark \\
		T11 \cite{VC-WNV-adapt} & Baseline & PPG $\to$ \textbf{LSTM} $\to$ STRAIGHT features & WaveNet & \cmark \\
		T16 \cite{cyclevae, vcc2020-baseline-cyclevae} & Baseline & WORLD features $\to$ \textbf{CycleVAE} $\to$ WORLD features & PWG & \xmark \\
		T22 \cite{vcc2020-asr-tts} & Baseline & Mel filterbanks $\to$ \textbf{seq2seq ASR} $\to$ text $\to$ \textbf{seq2seq TTS} $\to$ mel filterbanks & PWG & \xmark \\
		\midrule
		T23 task 1 & NU system & mel filterbanks $\to$ \textbf{seq2seq} $\to$ mel filterbanks & PWG & \xmark \\
		T23 task 2 & NU system & WORLD features $\to$ \textbf{CycleVAE} $\to$ WORLD features & WaveNet & \cmark \\
		\bottomrule
	\end{tabular}
	\label{tab:comparing-systems}
    \vspace*{-4.5mm}
\end{table*}

\begin{figure*}[t]
	\centering
	\begin{subfigure}{0.48\textwidth}
		\centering
  		\includegraphics[width=\linewidth]{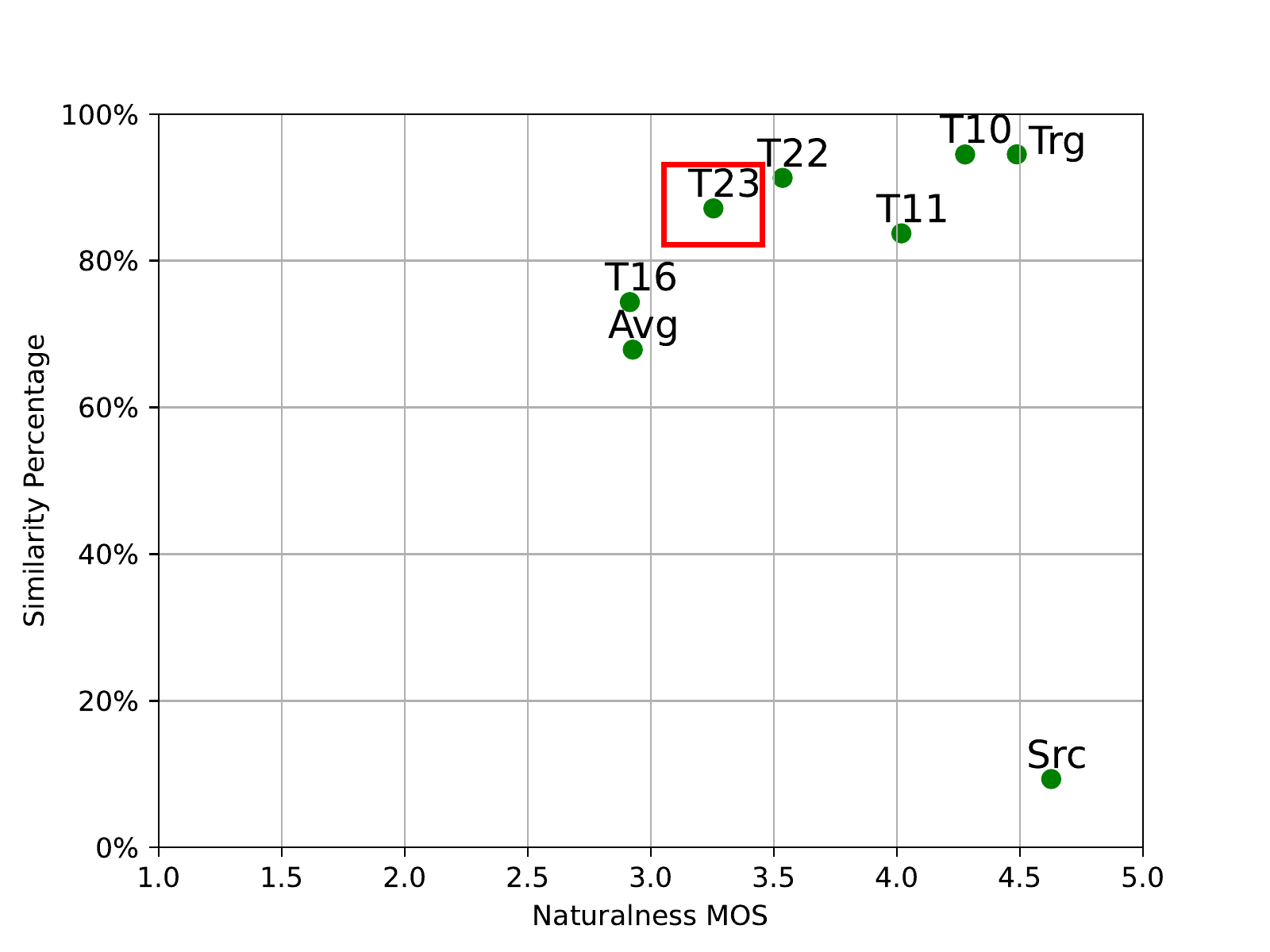}
        \caption{Task 1}
        \label{fig:task1-scatter}
	\end{subfigure}
	\begin{subfigure}{0.48\textwidth}
		\centering
  		\includegraphics[width=\linewidth]{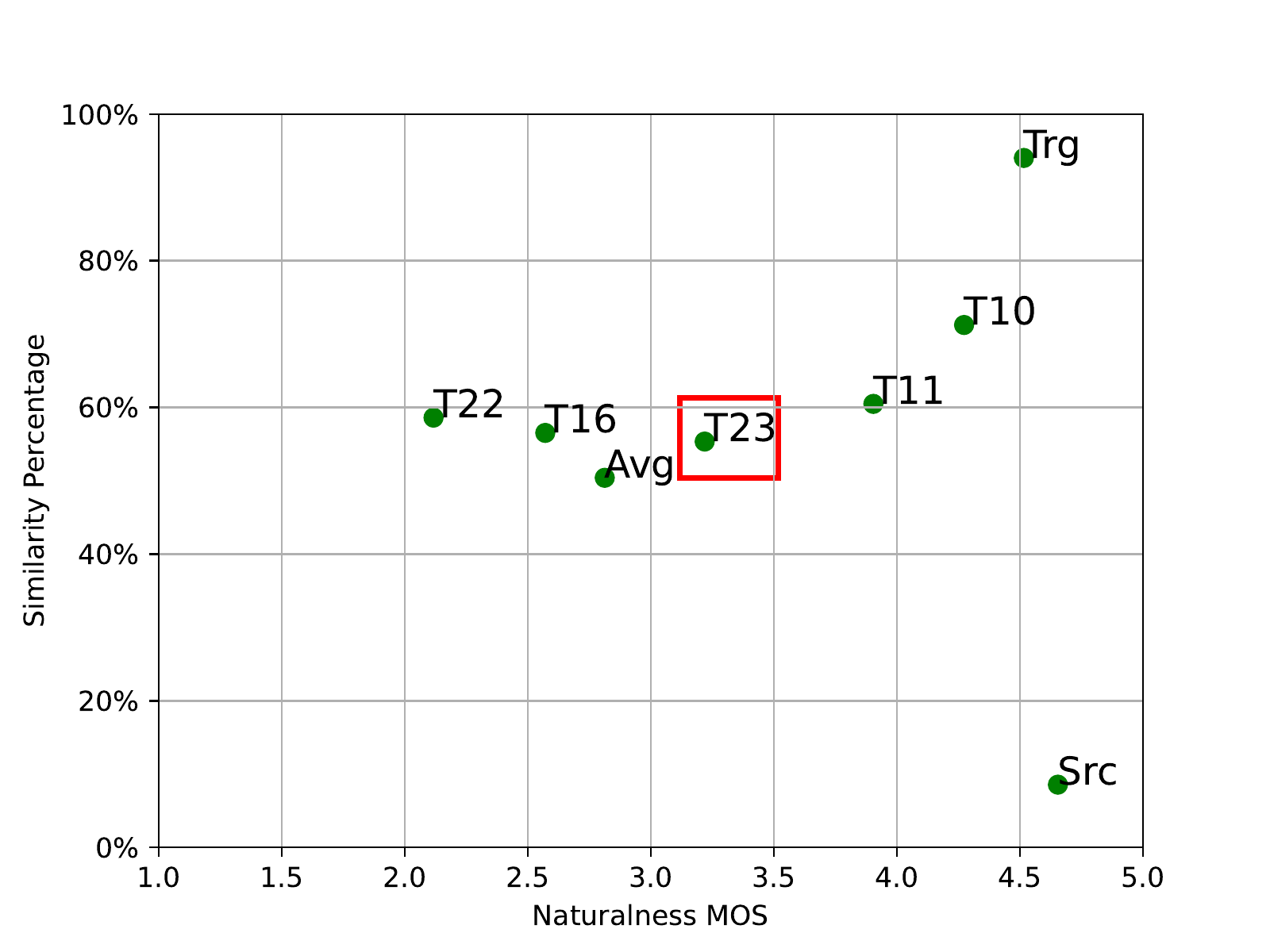}
        \caption{Task 2}
        \label{fig:task2-scatter}
	\end{subfigure}
	\centering
    \vspace*{-2.5mm}
	\caption{Scatter plot of naturalness and similarity scores. Our system, T23, is indicated by the red boxes.}
	\label{fig:results}
    \vspace*{-3.5mm}
\end{figure*}

\section{Evaluation Results}

The VCC2020 organizing committee conducted a large-scale subjective test on all submitted systems for both tasks 1 and 2. The evaluations included naturalness and similarity tests. We first describe the evaluation protocols. Then, we compare the performance of our entry with a subset of the participating systems, as briefly described in Table~\ref{tab:comparing-systems}. The results are shown in Figure~\ref{fig:results}.

\subsection{Evaluation protocol}

In the naturalness test, a five-point mean opinion score (MOS) test was adopted, where listeners were instructed to rate the naturalness of each speech clip from 1 to 5. In the similarity test, listeners were presented a converted utterance and a ground truth target utterance, and they were asked to determine whether the two utterances were spoken by the same person on a four-point scale. Figure~\ref{fig:results} shows the overall results\footnote{Although the official listening report contained results from Japanese and English listeners, we only report results of English listeners since the two listener groups shared a similar tendency of their results.}.

\vspace*{-0.5mm}
\subsection{Task 1 results}
\vspace*{-1mm}
Figure~\ref{fig:task1-scatter} shows the scatter plot of the naturalness and similarity results of task 1. First, our system showed better results than the average of all submitted systems in terms of both naturalness and similarity. Compared with the T16 baseline, which also used the non-AR PWG, our system was superior in terms of both naturalness and similarity. As T16 used a frame-based method (Cycle-VAE), this result demonstrates the effectiveness of seq2seq modeling. Next, without using the AR WaveNet vocoder as T16 did, our system still performed better in terms of similarity, again showing the power of seq2seq models when it comes to modeling speaker identity. Finally, our system could not outperform the last baseline, T22, which also used PWG and seq2seq conversion models. This suggests that parallel acoustic feature mapping may not be the best approach when it comes to seq2seq VC modeling, although an integrated model should theoretically outperform a cascade model such as T22. Analysis on this result will be an important future work.

\vspace*{-0.5mm}
\subsection{Task 2 results}
\vspace*{-1mm}
The result of task 2 is shown in Figure~\ref{fig:task2-scatter}. It can be observed that our system could yield significant naturalness improvements compared with the T16 baseline, which used CycleVAE-based spectral modeling but with a non-AR neural vocoder, i.e, Parallel WaveGAN. Even though the condition of CycleVAE training was not exactly the same, where our system used more data than the T16 baseline, the quality of speech waveforms is heavily affected by the use of the neural vocoder. This was also observed in our internal assessment, where we found that the quality of our system and that of the T16 baseline when using a conventional vocoder, such as WORLD, was similar. Thus, further improvements of feature mapping accuracy, to improve not only naturalness but also speaker similarity, can be achieved by using more constraints for speaker-independent space, such as PPG-based VC, where all of the best systems in VCC 2020 used this approach, as well as baseline T11 and the winner T10.

\vspace*{-1mm}
\section{Conclusions}
\vspace*{-0.5mm}
In this paper, we described the NU VC systems for the VCC2020. The task 1 system was based on VTN, a seq2seq VC model, with the help of synthetic parallel data generated with a TTS system. The task 2 system combined CycleVAE, a frame-based spectral mapping model, with WaveNet, an AR neural vocoder. By comparing with the baseline systems, we confirmed that seq2seq modeling and AR neural vocoders can improve conversion similarity and naturalness of the converted speech, respectively. For future work, we will investigate various aspects of the synthetic parallel data method, such as how the choice, size, and quality of the synthetic data affect the conversion performance. Also, the combination of seq2seq VC models and AR vocoders would be another promising direction.

\vspace*{-1mm}
\section{Acknowledgements}
\vspace*{-0.5mm}
This work was partly supported by JST, CREST Grant Number JPMJCR19A3,
and JSPS KAKENHI Grant Number 17H06101.

\bibliographystyle{IEEEtran}

\bibliography{mybib}

% Generated by IEEEtran.bst, version: 1.13 (2008/09/30)
\begin{thebibliography}{10}
\providecommand{\url}[1]{#1}
\csname url@samestyle\endcsname
\providecommand{\newblock}{\relax}
\providecommand{\bibinfo}[2]{#2}
\providecommand{\BIBentrySTDinterwordspacing}{\spaceskip=0pt\relax}
\providecommand{\BIBentryALTinterwordstretchfactor}{4}
\providecommand{\BIBentryALTinterwordspacing}{\spaceskip=\fontdimen2\font plus
\BIBentryALTinterwordstretchfactor\fontdimen3\font minus
  \fontdimen4\font\relax}
\providecommand{\BIBforeignlanguage}[2]{{%
\expandafter\ifx\csname l@#1\endcsname\relax
\typeout{** WARNING: IEEEtran.bst: No hyphenation pattern has been}%
\typeout{** loaded for the language `#1'. Using the pattern for}%
\typeout{** the default language instead.}%
\else
\language=\csname l@#1\endcsname
\fi
#2}}
\providecommand{\BIBdecl}{\relax}
\BIBdecl

\bibitem{VC}
Y.~Stylianou, O.~Cappe, and E.~Moulines, ``{Continuous probabilistic transform
  for voice conversion},'' \emph{IEEE Transactions on Speech and Audio
  Processing}, vol.~6, no.~2, pp. 131--142, 1998.

\bibitem{GMM-VC}
T.~Toda, A.~W. Black, and K.~Tokuda, ``{Voice Conversion Based on
  Maximum-Likelihood Estimation of Spectral Parameter Trajectory},'' \emph{IEEE
  Transactions on Audio, Speech, and Language Processing}, vol.~15, no.~8, pp.
  2222--2235, 2007.

\bibitem{vcc2020summary}
Y.~Zhao, W.-C. Huang, X.~Tian, J.~Yamagishi, R.~K. Das, T.~Toda, T.~Kinnunen,
  and Z.~Ling, ``Voice conversion challenge 2020 — intra-lingual semiparallel
  and cross-lingual voice conversion —,'' in \emph{ISCA Joint Workshop for
  the Blizzard Challenge and Voice Conversion Challenge 2020}, 2020.

\bibitem{S2S}
I.~Sutskever, O.~Vinyals, and Q.~V. Le, ``{Sequence to Sequence Learning with
  Neural Networks},'' in \emph{Advances in Neural Information Processing
  Systems}, 2014, pp. 3104--3112.

\bibitem{S2S-NMT-Bah}
D.~Bahdanau, K.~Cho, and Y.~Bengio, ``{Neural machine translation by jointly
  learning to align and translate},'' \emph{arXiv preprint arXiv:1409.0473},
  2014.

\bibitem{S2S-NMT-Luong}
T.~Luong, H.~Pham, and C.~D. Manning, ``Effective approaches to attention-based
  neural machine translation,'' in \emph{Proc. of the Conference on Empirical
  Methods in Natural Language Processing}, Lisbon, Portugal, Sep. 2015, pp.
  1412--1421.

\bibitem{ATT-S2S-VC}
K.~{Tanaka}, H.~{Kameoka}, T.~{Kaneko}, and N.~{Hojo}, ``{ATTS2S-VC:
  Sequence-to-sequence Voice Conversion with Attention and Context Preservation
  Mechanisms},'' in \emph{IEEE International Conference on Acoustics, Speech
  and Signal Processing (ICASSP)}, May 2019, pp. 6805--6809.

\bibitem{S2S-iFLYTEK-VC}
J.~{Zhang}, Z.~{Ling}, L.~{Liu}, Y.~{Jiang}, and L.~{Dai},
  ``{Sequence-to-Sequence Acoustic Modeling for Voice Conversion},''
  \emph{IEEE/ACM Transactions on Audio, Speech, and Language Processing},
  vol.~27, no.~3, pp. 631--644, 2019.

\bibitem{wnv}
A.~van~den Oord, S.~Dieleman, H.~Zen, K.~Simonyan, O.~Vinyals, A.~Graves,
  N.~Kalchbrenner, A.~W. Senior, and K.~Kavukcuoglu, ``{WaveNet}: {A}
  generative model for raw audio,'' \emph{CoRR arXiv preprint
  arXiv:1609.03499}, 2016.

\bibitem{wnv-voc}
A.~Tamamori, T.~Hayashi, K.~Kobayashi, K.~Takeda, and T.~Toda,
  ``Speaker-dependent {W}ave{N}et vocoder,'' in \emph{Proc. INTERSPEECH}, 2017,
  pp. 1118--1122.

\bibitem{shallow-wnv}
P.~L. Tobing, Y.-C. Wu, T.~Hayashi, K.~Kobayashi, and T.~Toda, ``Efficient
  shallow {WaveNet} vocoder using multiple samples output based on {L}aplacian
  distribution and linear prediction,'' in \emph{Proc. ICASSP}, 2020, pp.
  7204--7208.

\bibitem{samplernn}
S.~Mehri, K.~Kumar, I.~Gulrajani, R.~Kumar, S.~Jain, J.~Sotelo, A.~Courville,
  and Y.~Bengio, ``Sample{RNN}: {A}n unconditional end-to-end neural audio
  generation model,'' in \emph{Proc. ICLR}, Apr. 2017.

\bibitem{wavernn}
N.~Kalchbrenner, E.~Elsen, K.~Simonyan, S.~Noury, N.~Casagrande, E.~Lockhart,
  F.~Stimberg, A.~van~den Oord, S.~Dieleman, and K.~Kavukcuoglu, ``Efficient
  neural audio synthesis,'' in \emph{Proc. ICML}, July 2018, pp. 2415--2424.

\bibitem{lpcnet}
J.-M. Valin and J.~Skoglund, ``{LPCN}et: Improving neural speech synthesis
  through linear prediction,'' in \emph{Proc. ICASSP}, May 2019, pp.
  5891--5895.

\bibitem{clarinet}
W.~Ping, K.~Peng, and J.~Chen, ``Clari{N}et: Parallel wave generation in
  end-to-end text-to-speech,'' in \emph{Proc. ICLR}, May 2019.

\bibitem{waveglow}
R.~Prenger, R.~Valle, and B.~Catanzaro, ``Wave{G}low: A flow-based generative
  network for speech synthesis,'' in \emph{Proc. ICASSP}, May 2019, pp.
  3617--3621.

\bibitem{flowavenet}
S.~Kim, S.-G. Lee, J.~Song, J.~Kim, and S.~Yoon, ``{F}lo{W}ave{N}et: A
  generative flow for raw audio,'' in \emph{Proc. ICML}, June 2019, pp.
  3370--3378.

\bibitem{nsf_2019}
X.~Wang, S.~Takaki, and J.~Yamagishi, ``Neural source-filter-based waveform
  model for statistical parametric speech synthesis,'' in \emph{Proc. ICASSP},
  May 2019, pp. 5916--5920.

\bibitem{parallel-wavegan}
R.~{Yamamoto}, E.~{Song}, and J.~{Kim}, ``{Parallel WaveGAN: A Fast Waveform
  Generation Model Based on Generative Adversarial Networks with
  Multi-Resolution Spectrogram},'' in \emph{IEEE International Conference on
  Acoustics, Speech and Signal Processing (ICASSP)}, 2020, pp. 6199--6203.

\bibitem{melgan}
K.~Kumar, R.~Kumar, T.~de~Boissiere, L.~Gestin, W.~Z. Teoh, J.~Sotelo,
  A.~de~Br\'{e}bisson, Y.~Bengio, and A.~C. Courville, ``Mel{GAN}: Generative
  adversarial networks for conditional waveform synthesis,'' in \emph{Proc.
  NeurIPS}, Dec. 2019, pp. 14\,910--14\,921.

\bibitem{gantts}
M.~Bi{\'n}kowski, J.~Donahue, S.~Dieleman, A.~Clark, E.~Elsen, N.~Casagrande,
  L.~C. Cobo, and K.~Simonyan, ``High fidelity speech synthesis with
  adversarial networks,'' in \emph{Proc. ICLR}, Apr. 2020.

\bibitem{qppwg}
Y.-C. Wu, T.~Hayashi, T.~Okamoto, H.~Kawai, and T.~Toda, ``Quasi-periodic
  parallel wavegan vocoder: A non-autoregressive pitch-dependent dilated
  convolution model for parametric speech generation,'' in \emph{Proc.
  Interspeech}, Oct. 2020.

\bibitem{straight}
H.~Kawahara, I.~Masuda-Katsuse, and A.~De~Cheveigne, ``Restructuring speech
  representations using a pitch-adaptive time--frequency smoothing and an
  instantaneous-frequency-based {F0} extraction: Possible role of a repetitive
  structure in sounds,'' \emph{Speech Communication}, vol.~27, no. 3-4, pp.
  187--207, 1999.

\bibitem{world}
M.~Morise, F.~Yokomori, and K.~Ozawa, ``{WORLD}: a vocoder-based high-quality
  speech synthesis system for real-time applications,'' \emph{IEICE
  Transactions on Information and Systems}, vol.~99, no.~7, pp. 1877--1884,
  2016.

\bibitem{VTN}
W.-C. Huang, T.~Hayashi, Y.-C. Wu, H.~Kameoka, and T.~Toda, ``Voice transformer
  network: Sequence-to-sequence voice conversion using transformer with
  text-to-speech pretraining,'' \emph{arXiv preprint arXiv:1912.06813}, 2019,
  to appear in Interspeech 2020.

\bibitem{VTN-TASLP}
------, ``Pretraining techniques for sequence-to-sequence voice conversion,''
  \emph{arXiv preprint arXiv:2008.03088}, 2020.

\bibitem{cyclevae}
P.~L. Tobing, Y.-C. Wu, T.~Hayashi, K.~Kobayashi, and T.~Toda, ``{Non-Parallel
  Voice Conversion with Cyclic Variational Autoencoder},'' in \emph{Proc.
  Interspeech}, 2019, pp. 674--678.

\bibitem{S2S-parrotron-VC}
F.~Biadsy, R.~J. Weiss, P.~J. Moreno, D.~Kanvesky, and Y.~Jia, ``{Parrotron: An
  End-to-End Speech-to-Speech Conversion Model and its Applications to
  Hearing-Impaired Speech and Speech Separation},'' in \emph{Proc.
  Interspeech}, 2019, pp. 4115--4119.

\bibitem{CMU-Arctic}
J.~Kominek and A.~W. Black, ``{The CMU ARCTIC speech databases},'' in
  \emph{Fifth ISCA Workshop on Speech Synthesis}, 2004.

\bibitem{vcc2020-baseline-cyclevae}
P.~L. Tobing, Y.-C. Wu, and T.~Toda, ``The baseline system of voice conversion
  challenge 2020 with cyclic variational autoencoder and parallel wavegan,'' in
  \emph{ISCA Joint Workshop for the Blizzard Challenge and Voice Conversion
  Challenge 2020}, 2020.

\bibitem{vae}
D.~P. Kingma and J.~Ba, ``Auto-encoding variational bayes,'' \emph{arXiv
  preprint arXiv:1312.6114}, 2013.

\bibitem{vaevc}
C.-C. Hsu, H.-T. Hwang, Y.-C. Wu, Y.~Tsao, and H.-M. Wang, ``Voice conversion
  from non-parallel corpora using variational auto-encoder,'' in \emph{Proc.
  APSIPA}, 2016, pp. 1--6.

\bibitem{vc-wnv-ft}
P.~L. Tobing, Y.-C. Wu, T.~Hayashi, K.~Kobayashi, and T.~Toda, ``Voice
  conversion with {CycleRNN}-based spectral mapping and finely tuned {WaveNet}
  vocoder,'' \emph{IEEE Access}, vol.~7, pp. 171\,114--171\,125, 2019.

\bibitem{vctk}
C.~Veaux, J.~Yamagishi, and K.~MacDonald, ``{CSTR} {VCTK} corpus: English
  multi-speaker corpus for {CSTR} voice cloning toolkit,'' 2017.

\bibitem{wsola-f0}
K.~Kobayashi, T.~Toda, and S.~Nakamura, ``F0 transformation techniques for
  statistical voice conversion with direct waveform modification with spectral
  differential,'' in \emph{Proc. SLT}, 2016, pp. 693--700.

\bibitem{mel-cep}
K.~Tokuda, T.~Kobayashi, T.~Masuko, and S.~Imai, ``Mel-generalized cepstral
  analysis - a unified approach to speech spectral estimation,'' in \emph{Proc.
  ICSLP}, Yokohama, Japan, Sep. 1994, pp. 1043--1046.

\bibitem{VC-WNV-adapt}
L.-J. Liu, Z.-H. Ling, Y.~Jiang, M.~Zhou, and L.-R. Dai, ``{WaveNet Vocoder
  with Limited Training Data for Voice Conversion},'' in \emph{Proc.
  Interspeech}, 2018, pp. 1983--1987.

\bibitem{vcc2020-asr-tts}
W.-C. Huang, T.~Hayashi, S.~Watanabe, and T.~Toda, ``The sequence-to-sequence
  baseline for the voice conversion challenge 2020: Cascading asr and tts,'' in
  \emph{ISCA Joint Workshop for the Blizzard Challenge and Voice Conversion
  Challenge 2020}, 2020.

\end{thebibliography}

\end{document}